\documentclass{book}
\usepackage[T1]{fontenc}
\usepackage[utf8]{inputenc}
\usepackage{graphicx}
\usepackage{latexsym}
\usepackage{amsfonts}
\usepackage{amssymb}
\usepackage{amsbsy}
\usepackage{color}
\usepackage{amsmath}
\usepackage{multicol}
\usepackage{float}
\usepackage{subfigure}
\usepackage{algorithm}
\usepackage{algorithmic}
\usepackage[dvips]{epsfig}

\usepackage{caption}
\usepackage{enumitem}
\usepackage{graphicx}
\usepackage{latexsym}
\usepackage{amsfonts}
\usepackage{amssymb}
\usepackage{amsbsy}
\usepackage{color}
\usepackage{amsmath}
\usepackage{multicol}
\usepackage{float}
\usepackage{subfigure}
\usepackage{algorithm}
\usepackage{algorithmic}
\usepackage[dvips]{epsfig}

\newcommand{\chapternote}[1]{%
 \let\thempfn\relax
  \footnotetext[0]{\emph{#1}}
  }
\DeclareMathOperator*{\arginf}{\arg\inf}

\title{Mean-Field Game-Theoretic Edge Caching}

\begin{document}

\chapter{Mean-Field Game-Theoretic Edge Caching}

Hyesung Kim\footnote{H. Kim was with the School of Electrical and Electronic Engineering, Yonsei University, and is currently with Samsung Research, 135-090 Seoul, Korea (email: hye1207@gmail.com).}, 
Jihong Park\footnote{J. Park is with the School of Information Technology, Deakin University, Geelong, VIC 3220, Australia (email: jihong.park@deakin.edu.au).},\\
Mehdi Bennis\footnote{M. Bennis is with the Centre of Wireless Communications, University of Oulu, 90014 Oulu, Finland (email: mehdi.bennis@oulu.fi).},
Seong-Lyun Kim\footnote{S.-L. Kim is with the School of Electrical and Electronic Engineering, Yonsei University, 120-749 Seoul, Korea (email: slkim@yonsei.ac.kr).},
and M\'erouane Debbah\footnote{M. Debbah is with Université Paris-Saclay, CNRS, CentraleSupélec, 91190 Gif-sur-Yvette, France (e-mail:
merouane.debbah@centralesupelec.fr) and the Lagrange Mathematical and
Computing Research Center, 75007 Paris, France.}

\section{Introduction}

Mobile networks are envisaged to be extremely densified in 5G and beyond to cope with the ever-growing user demand \cite{UDN_sur, JensUdnMag,Shim16,PetarURLLC:17}. Edge caching is a key enabler of such an ultra-dense network (UDN), through which popular content is prefetched at each small base station (SBS) and downloaded with low latency \cite{Living,caching_tradeoff:2016} while alleviating the significant backhaul congestion between a data server and a large number of SBSs \cite{caching_effect}. Focusing on this, in this chapter we study the content caching strategy of an \emph{ultra-dense edge caching network (UDCN)}. Optimizing the content caching of a UDCN is a crucial yet challenging problem. Due to the sheer amount of SBSs, even a small misprediction of user demand may result in a large amount of useless data cached in capacity-limited storages. Furthermore, the variance of interference is high due to short inter-SBS distances~\cite{interf_udn}, making it difficult to evaluate cached data downloading rates, which is essential in optimizing the caching file sizes. To resolve these problems, we first present a spatio-temporal user demand model in continuous time, in which the long-term and short-term content popularity variations at a specific location are modeled using the Chinese restaurant process (CRP) and the Ornstein-Uhlenbeck (OU) process, respectively. Based on this, we aim to develop a scalable and distributed edge caching algorithm by leveraging the mean-field game (MFG) theory~\cite{MF_Caine1,MF_Caines2}.

To this end, at first the problem of optimizing distributed edge caching strategies in continuous time is cast as a non-cooperative stochastic differential game (SDG). As the game player, each SBS decides how much portion of each content file is prefetched by minimizing its long run average (LRA) cost that depends on the prefetching overhead, cached file downloading rates under inter-SBS interference, and overlapping cached files among neighboring SBSs, i.e., content overlap. This minimization problem is tantamount to solving a partial differential equation (PDE) called the Hamilton-Jacobi-Bellman equation (HJB) \cite{exist_HJBsol1}. The major difficulty is that the HJB solution of an SBS is intertwined with the HJB solutions of other SBSs, as they interact with each other through the inter-SBS interference and content overlap. The complexity of this problem is thus increasing exponentially with the number of SBSs, which is unfit for a UDCN. Alternatively, exploiting MFG, we decouple the SBS interactions in a way that each SBS interacts only with a virtual agent whose behaviors follow the state distribution of the entire SBS population, known as mean-field (MF) distribution. For the given SBS population, the MF distribution is uniquely and derived by locally solving a PDE, called the Fokker-Planck-Kolmogorov equation (FPK). Consequently, the optimal caching problem at each SBS boils down to solving a single pair of HJB and FPK, regardless of the number of SBSs. Such an MF approximation guarantees achieving the epsilon Nash equilibrium \cite{MF_Caine1,EV_JSAC}, when the number of SBSs is very large (theoretically approaching infinity) while their aggregate interactions are bounded. Both conditions are satisfied in a UDCN~\cite{ICC,TVT_MFCaching}, mandating the use of MFG.

To describe the MFG-theoretic caching framework and show its effectiveness for a UDCN, this chapter is structured as follows. Related works on UDCN analysis and MFG-theoretic approaches are briefly reviewed in chapter~\ref{Related work}. The network, channel, and caching models as well as the spatio-temporal dynamics of user demand, caching, and interference are described in chapter~\ref{C_system_model}. For the SBS supporting a reference user, its optimal caching problem is formulated and solved using MFG in chapter \ref{C_formulation}. The performance of the MFG-theoretic edge caching is numerically evaluated in terms of LRA and the content overlap amount in chapter \ref{C_numerical}, followed by concluding remarks in chapter \ref{Conclusion_remark}.

\section{Related Works} \label{Related work}
Edge caching in cellular networks has received significant attention in 5G and beyond \cite{caching_tradeoff:2016,caching_effect,PIEEE1}. In the context of MFG-theoretic edge caching in a UDN, we briefly review its preceding user demand models, interference analysis, and MFG-based applications as follows.


\textbf{User Demand Model and Interference Analysis.}\quad The effectiveness of edge caching is affected significantly by user demand according to content popularity. The user demand model in most of the works on edge caching relies commonly on the Zipf's law. Under this model, the content popularity in the entire network region is static and follows a truncated power law~\cite{Zipf_1}, which is too coarse to reflect spatio-temporal content popularity dynamics in a UDCN. A time-varying user demand model has been considered in \cite{caching_dyna1,MFG_caching} while ignoring spatial characteristics, which motivated us to seek for a more detailed user demand model reflecting spatio-temporal content popularity variations.

The spatial characteristics of interference dynamics has been analyzed in \cite{caching_dyna2,spatially_corr} using stochastic geometry. These works however rely on a globally static user demand model, and thus ignore the temporal and local dynamics of interference \cite{local_global}. By contrast, in this chapter we consider the spatio-temporal content popularity dynamics, and analyze their impact on interference.

The impacts of SBS densification on interference in a UDN have been investigated in \cite{interf_udn,JHP1,JHP2,PIMRC17,JParkSpaSWiN:17,KimAccess,Nemati}, in which the interference dynamics is governed by the spatial dynamics of user demand, i.e., locations \cite{interf_udn}. While interesting, these works neglect temporal user demand variations. It is worth noting that a recent study \cite{ICC} has considered spatio-temporal user demand fluctuations. However, it does not take into account temporal content popularity correlation. The gap has been filled by its follow-up work \cite{TVT} that models the correlated content popularity using the CRP, which is addressed in this chapter.

\textbf{MFG Applications.}\quad
The MFG theory is built upon an asymptotically large number of agents in a non-cooperative game. This fits naturally with a UDN within which assuming an infinite number of SBSs becomes reasonable \cite{UDN_sur,JensUdnMag,interf_udn,PIEEE2}. In this respect, SBS transmit power control and the user mobility management in a UDN have been studied in \cite{JHP1,JHP2}. For a massive number of drones, their rate-maximizing altitude control and collision-avoid path planning have been investigated in \cite{Alt} and \cite{SPAWC_hs,Hamid1, Hamid2,park2020extreme}, respectively. In a similar vein, in this chapter we utilize the MFG theory to simplify the spatio-temporal analysis on interference and content overlap in a UDCN. One major limitation of MFG-based methods is that solving a pair of HJB-FPK PDEs may still be challenging when the agent's state dimension is large. In fact, existing PDE solvers rely mostly on the Euler discretization method such that the derivatives in a PDE are approximated using finite differences. To guarantee the convergence of a numerical PDE solution, the larger state dimension is considered, the finer discretization step size is required under the Courant-Friedrichs-Lewy (CFL) condition \cite{CFL}, increasing the computing complexity. To avoid this problem, {in this chapter we describe the state of each SBS separately for each content file, reducing the dimension of each PDE.} Alternatively, machine learning methods have been applied in recent works \cite{Hamid1,Hamid2,Hamid3} by which solving a PDE is recast as a regression learning problem. By leveraging this method, incorporating large-sized edge caching states (e.g., joint optimization of transmit power, caching strategy, and mobility management) could be an interesting topic for future research.

\section{System Model}
\label{C_system_model}

\subsection{Network, Channel, and Caching Model}
In this section, we describe a downlink UDN under study, followed by its communication channel and caching models.

\textbf{Network Model.}\quad SBSs and their users are independently and uniformly distributed in a two-dimensional plane $\mathbb{R}^2$ with finite densities, forming two independent Poisson point processes (PPPs)~\cite{interf_udn,interf_var}. Following \cite{interf_udn}, the network is assumed to be a UDN such that SBS density $\lambda_b$ is much higher than user density $\lambda_u$, i.e., $\lambda_b \gg \lambda_u$. In this UDN, the $i$-th user is located at the coordinates $y_i\in\mathbb{R}^2$, and receives signals from multiple SBSs located within a reception ball $b(y_i,R)$ centered at $y_i$ with radius $R>0$, as depicted in Fig.~1. The radius $R$ can be determined based on the noise floor so that the average received signal power should be larger than the noise floor. When $R\rightarrow \infty$, the reception ball model becomes identical to a conventional PPP based network model in stochastic geometric analysis \cite{interf_var,finitePPP}.

\textbf{Channel and Antenna Pattern Models.}\quad
The transmitted signals from SBSs experience path-loss attenuation and multi-path fading. Specifically, the path loss $l_{k,i}$ from the $k$-th SBS located at $z_k\in\mathbb{R}^2$ to the $i$-th user at $y_i\in\mathbb{R}^2$ is given as $l_{k,i}=\min(1,||z_k-y_i||^{-\alpha})$, where $\alpha>2$ is the path-loss exponent. The transmitted signals experience an independent and identically distributed fading with the coefficient $g_{k,i}(t)$. We assume that the fading coefficient is not temporally correlated. Consequently, the received signal power at the $i$-th user is given as $S(t)=P|h_{k,i}(t)|^2$, where $P$ denotes the transmit power of every SBS, and $h_{k,i}$ is the channel gain determined by $|h_{k,i}(t)|^2=l_{k,i}|g_{k,i}(t)|^2$. Next, the transmission of each SBS is directional using $N_a$ antennas. Following \cite{MIMO}, the beam pattern follows a sectored uniform linear array model, in which the center of the mainlobe beam points at the receiving user. The mainlobe gain is given as $N_a$ with the beam width $\theta_{N_a} = 2\pi/\sqrt{N_a}$ while ignoring side lobes.

\textbf{Caching Model.}\quad Consider a set $\mathcal{M}$ of $M$ content files in total, each of which is encoded using the maximum distance separable dateless code \cite{MDS}. At time $t$, a fraction $p_{k,j}(t)\in[0,1]$ of the $j$-th content file with the size $L_j$ is prefetched to the $k$-th SBS from a remote server through a capacity-limited backhaul link as shown in Fig. \ref{system_model_CRP}. The SBS is equipped with a data storage of size $C_{k,j}$ assigned for the content file $j$, and therefore we have $p_{k,j}(t) L_j \leq C_{k,j}$. Each user in the network requests the $j$-th content file with probability $x_j$. Within the user's content request range $R_c>0$, there exists a set $\mathcal{N}$ of $N$ SBSs~\cite{interf_udn}. If multiple SBSs cached the requested file (i.e., cache hitting), then the user downloads the file from a randomly selected SBS. If there is no SBS cached the requested file (i.e., cache missing), then the file is downloaded to a randomly selected SBS from the remote server via the backhaul, which is then delivered to the user from the SBS. At time $t$, the goal of the $k$-th SBS is to determine its file caching fraction vector $\boldsymbol{p}_k(t)=\{p_{k,1}(t),\cdots,p_{k,j}(t),\cdots p_{k,M}(t)\}$.

\subsection{Spatio-Temporal Dynamics of Demand, Caching, and Interference} \label{C-st dynamics}

The effectiveness of caching strategies is affected by spatio-temporal dynamics of content popularity among users, backhaul and storage capacities in SBSs, and interference across SBSs, as we elaborate next.

\textbf{User Demand Dynamics.}
The user demand on content files is often modeled as a Zipf distribution \cite{Zipf_1}. Such a long-term user demand pattern in a wide area is too coarse to capture the spatial demand and its temporal variations \cite{local_global}, calling for a detailed spatio-temporal user demand model for a UDCN. To this end, we consider that each SBS regularly probes the content popularity within the distance $R_s$, and the content popularity for each SBS follows an independent stochastic process. For the content popularity of each SBS, its temporal dynamics is described by the long-term fluctuations across time $t=T, 2T, \cdots ,\kappa T$ and short-term fluctuations over $t\in[(\kappa-1)T,\kappa T]$ as considered in \cite{MR_model}. These long-term and short-term content popularity dynamics are modeled using the Chinese restaurant process (CRP) and the Ornstein-Uhlenbeck (OU) process, respectively, as detailed next.

Following the CRP \cite{Living,CRP_1}, the long-term content popularity variations are described by the analogy of the table selection process in a Chinese restaurant. Here, a UDCN becomes a Chinese restaurant, wherein the content files and users are the tables and customers in the restaurant, respectively. Treating the restaurant table seating problem as a long-term content popularity updating model, we categorize content files into two groups: the set $U^r_k(\kappa T)$ of the files that have been requested by $N_k(\kappa T)$ users at least once at SBS~$k$ until time $\kappa T$; and the set $U^u_k(\kappa T)$ of the files that have not yet been requested until then. For these two groups, the mean popularity $\mu_{k,j}(\kappa T)$ of the $j$-th content file at SBS~$k$ during $t\in [(\kappa-1)T, \kappa T]$ is given~by:
\begin{align}
\mu_{k,j}(\kappa T)  = \left\{ \begin{array}{ll}
  \frac{n_{k,j}(\kappa T)-\nu}{N_k(\kappa T)+\theta} & \textrm{for $j \in U^r_k(\kappa T)$}\\
  \frac{\nu |U^r_k(\kappa T)|+\theta}{N_k(\kappa T)+\theta} & \textrm{for $j \in U^u_k(\kappa T)$},
\end{array} \right.  \label{avg_Nk}
\end{align}
where $n_{k,j}(\kappa T)$ is the number of accumulated downloading requests for the $j$-th content file at SBS $k$ until time $\kappa T$, and $\theta$ and $\nu$ are positive constants. Note that each content file popularity depends on the popularity of other files and the number of other files. Consequently, more popular files are more often requested, proportionally to the previous request history $n_{k,j}(\kappa T)$, while unpopular files can also be requested with a probability proportional to $\theta$ and $\nu$. For simplicity without loss of generality, we omit the index $\kappa T$ of $\mu_{k,j}(\kappa T)$, and focus only on the case when $\kappa=1$ hereafter.

Next, for a given mean content popularity $\mu_{k,j}$, at time $t$ during a short-term period $t \in [0 \leq t \leq T]$, the content request probability $x_{k,j}(t)$ of the $j$-th file at SBS $k$ is described by the OU process \cite{MR_model}, a stochastic differential equation (SDE) given as follows:
\begin{align}
  \text{d}x_{k,j}(t)= r(\mu_{k,j}-x_{k,j}(t))\text{d}t+\eta \text{d}W_{k,j}(t), \label{SDE_x}
  \end{align}   
  where $W_{k,j}(t)$ is the Wiener process, and $r$ and $\eta$ are positive constants. It describes that the short-term content popularity is drifted from the long-term mean content popularity $\mu_{k,j}$ by $x_{k,j}(t)$, and is randomly fluctuated by $W_{k,j}(t)$. 

\begin{figure}
    \centering
    \subfigure[Spatial popularity dynamics of the network at $t=0$.]{\includegraphics[width=.8\textwidth]{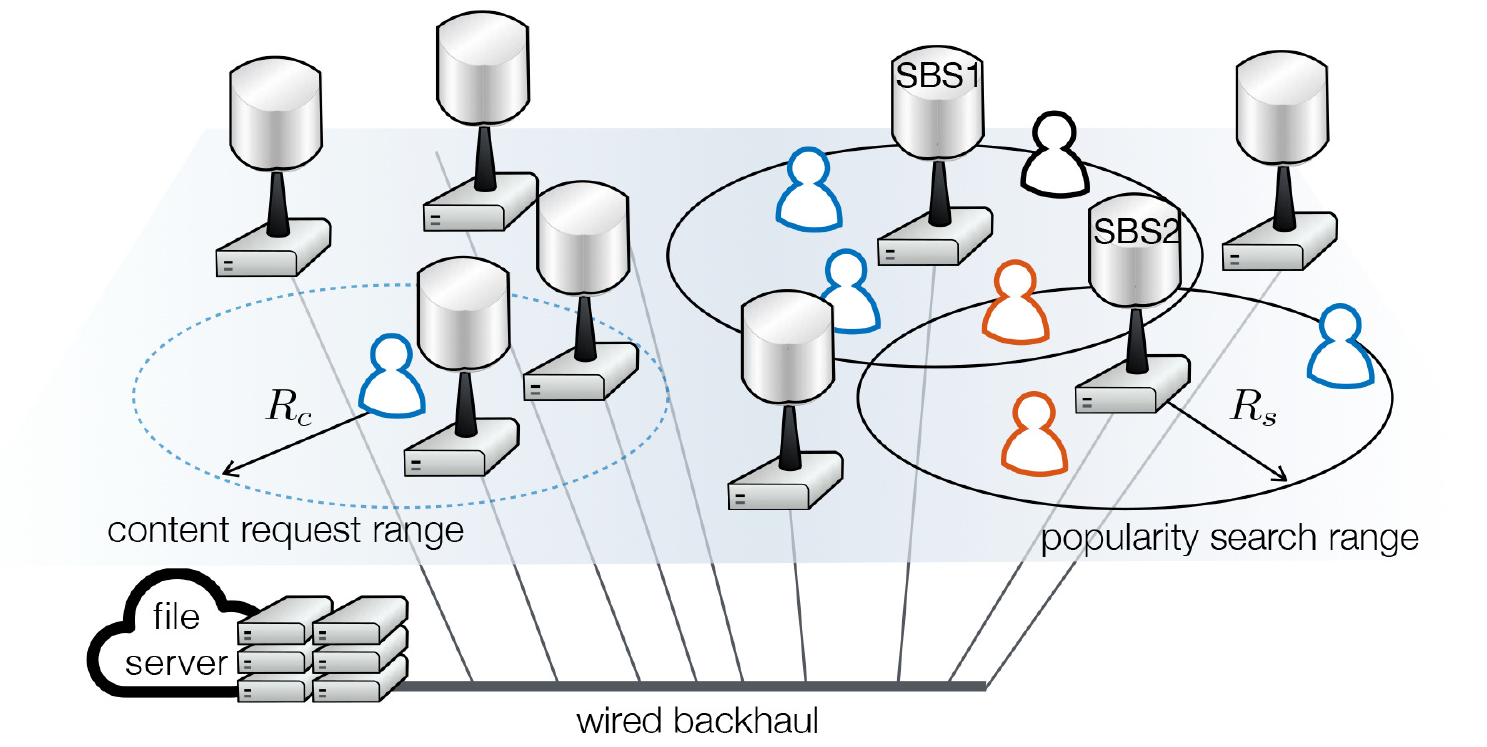}}
    \subfigure[Temporal popularity dynamics at SBSs 1 and 2 during $3T$.]{\includegraphics[width=.8\textwidth]{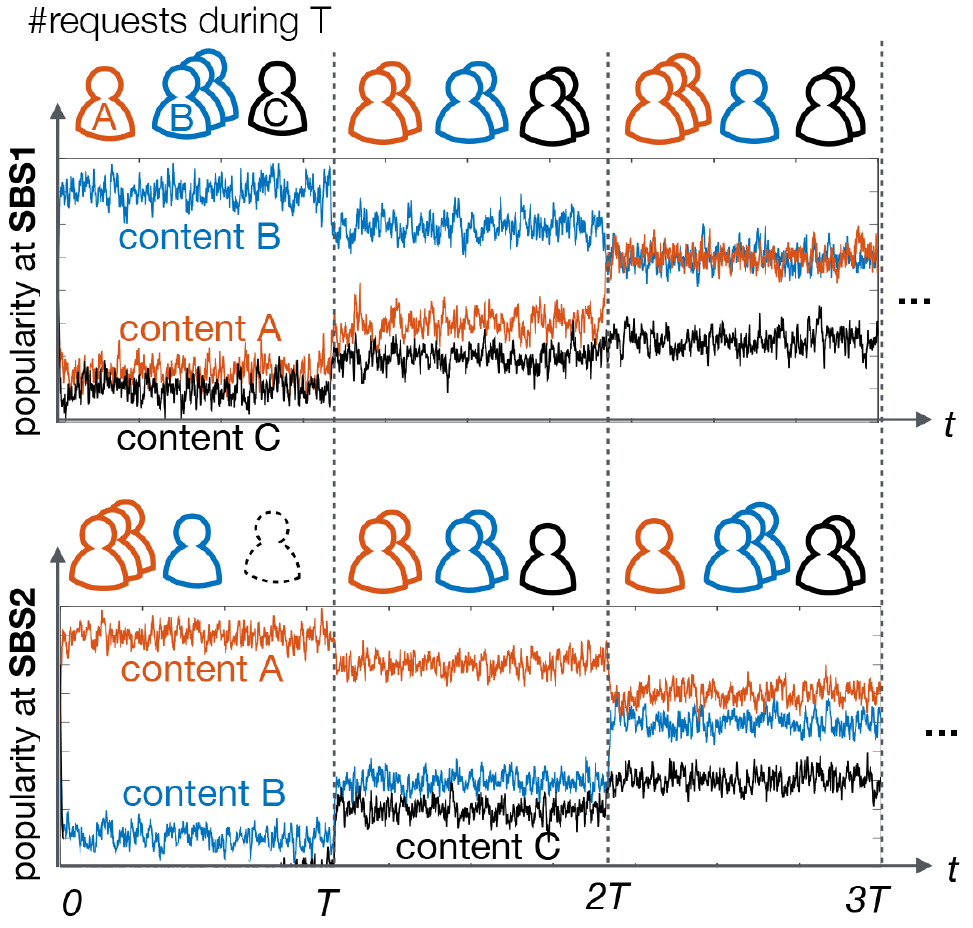}}
    \caption{\small{An illustration of a UDCN and its intrinsic spatio-temporal popularity dynamics. (a) Spatially dynamics of popularity  (b) Temporal dynamics where the content popularity changes for long-term and short-term duration. The long-term dynamics are captured by the Chinese restaurant process, which determines the mean popularity for a certain time period of $T$. During this period, the instantaneous popularity is captured by the mean reversion model following the OU process  \cite{MF_ref2}.  }}\label{system_model_CRP} 
    \end{figure}

Fig.~\ref{system_model_CRP} illustrates the user demand pattern generated from the aforementioned long-term and short-term content popularity dynamics. As observed by SBS 1 and SBS 2, for the same content files A, B, and C, these two spatially separated SBSs have different popularity dynamics, while at the same SBS each content popularity is updated according to a given temporal correlation. Furthermore, as shown in SBS 2 at around $t=T$, the previously unpopular file $C$ can emerge as an up-do-date popular~file.

    \textbf{Caching Dynamics.}\quad
    The remaining storage capacity varies according to the instantaneous caching strategy. 
    Let us assume that SBSs have finite storage size and discard content files at a rate of $e_{k,j}$ from the storage unit in order to make space for caching other contents.
    Considering the discarding rate, we model
    the evolution law of the storage unit at SBS $k$ as follows: 
    \begin{align}
    \text{d}Q_{k,j}(t)=(e_{k,j}-L_j p_{k,j}(t))\text{d}t, \label{SDE_q}
    \end{align}
    where $Q_{k,j}(t)$ denotes the remaining storage size dedicated to content $j$ of SBS $k$ at time $t$, and $L_j$ is data size of content~$j$. Note that $L_j p_{k,j}(t)$ represents the data size of content $j$ downloaded by SBS $k$ at time $t$. Since each user can download the requested file from one of multiple SBSs within its reception ball, for the given limited storage size, it is crucial to minimize overlapping content caching while maximizing the cache hitting rates, by determining the file caching fraction $p_{k,j}(t)$ at SBS $k$. This problem is intertwined with other SBSs' caching decisions, and the difficulty is aggravated under ultra-dense SBS deployment, seeking a novel solution with low-complexity using MFG to be elaborated in Sec.~\ref{C_formulation}.

    \textbf{Interference Dynamics}.\quad In a UDN, there is a considerable number of SBSs with no associated user within its coverage. These SBSs become idle and does not transmit any signal according to the definition of UDN ($\lambda_b\! \gg \!\lambda_u$) \cite{interf_udn}. Hence, this dormant SBS does not cause interference to neighbor SBSs. This leads to a spatially dynamic distribution of interference characterized by  users' locations. We assume that active SBSs have always data to transmit. Let us denote the SBS active probability by $p_a$. The aggregate interference is imposed by the active SBSs with probability $p_a$. Assuming that $p_a$ is homogeneous over SBSs yields $p_a \approx 1-[1+\lambda_u/(3.5\lambda_b)]^{-3.5}$ \cite{SMYu}. It provides that the density of interfering SBSs is equal to $p_a\lambda_b$. Then, at the typical user selected uniformly at random, the signal-to-interference-plus-noise (\textsf{SINR}) with $N_a$ number of transmit antennas is given as:
    \begin{align}
    \mathsf{SINR}(t) = \frac{N_a P |h(t)|^2}{{\sigma^2+\frac{\theta_{N_a}}{2\pi}N_aI^f(t)}}. \label{Eq:SINR}
    \end{align}
    where the aggregate interference $I^f(t)$ depends on the set $\Phi_R(p_a\lambda_b)$ of active SBS coordinates within the reception ball of radius $R$, given by $I^f(t)=\sum^{|\Phi_R(p_a\lambda_b)|}_k{ P|h_{k,i}(t)|^2}$. The term $\frac{\theta_{N_a}}{2\pi}N_a$ in \eqref{Eq:SINR} is given by the directional beam pattern. Given uniformly distributed (i.e., isotropically distributed) users, an SBS becomes an interferer with probability $\theta_{N_a}/(2\pi)$ with the main lobe gain $N_a$ and the beam width $\theta_{N_a} = 2\pi/\sqrt{N_a}$. Note that the interference term $I^f(t)$ depends on the spatial locations of SBSs and users (through $p_a$). This becomes a major bottleneck in anayzing UDCN, calling for a tractable way of handling $I^f(t)$ using MFG to be discussed in Sec.~\ref{C_formulation}.

    \section{Game theoretic formulation for edge caching}\label{C_formulation}

    We utilize the framework of non-cooperative games  to devise a fully distributed algorithm. The goal of each SBS $k$ is to determine its own caching amount ${p}_{k,j}^*(t)$ for content $j$ in order to minimize an LRA cost. 
    The LRA cost is determined by spatio-temporally varying content request probability,  network dynamics, content overlap, and aggregate inter-SBS interference. 
    As the SBSs' states and content popularity evolves, the caching strategies of the SBSs must adapt accordingly. {Minimizing the LRA cost under the spatio-temporal dynamics can be modeled as a dynamic stochastic differential game (SDG) \cite{MFG_application}.} 
    In the following subsection, we specify the impact of other SBSs' caching strategies and inter-SBS interference in the SDG by defining the LRA cost. 
    

    \subsection{Cost Functions}
    
    An instantaneous cost function $J_{k,j}(t)$ defines the LRA cost. It is affected by backhaul capacity, remaining storage size, average rate per unit bandwidth, and overlapping contents among SBSs. 
    SBS $k$ cannot download more than $B_{k,j}(t)$, defined as the allocated backhaul capacity for downloading content $j$ at time $t$. 
    In the proposed LRA cost, the download rate $L_jp_{k,j}(t)$ is prevented from exceeding the backhaul capacity constraint $B_{k,j}(t)$ by the backhaul cost function $\phi_{k,j}$ as $\phi_{k,j}(p_{k,j}(t))\!=\! -\log (B_{k,j}(t)-L_jp_{k,j}(t))$. If $L_jp_{k,j}(t)\! \geq \!{B_{k,j}(t)}$, the value of the cost function $\phi_{k,j}$ goes to infinity. This form of cost function is widely used to model barrier or constraint of available resources as in \cite{MFG_caching}.
    {As cached content files occupy the storage, it causes processing latency \cite{Storage_cost} or delay to search requested files by users.    
    This overhead cost is proportional to the cached data size in the storage unit.} To incorporate this, a storage cost function is proposed baed on the occupation ratio of the storage unit normalized by the storage size as follows:
    \begin{align}
    \psi_{k,j}(Q_{k,j}(t))= \gamma(C_{k,j}-Q_{k,j}(t))/{C_{k,j}},  \label{storage cost}
    \end{align}
    {where $Q_{k,j}(t)$ is storage cost function at time $t$, and $\gamma$ is a constant storage cost parameter.}
    \noindent Then, the global instantaneous cost is given by:
    \begin{align}
    J_{k,j}(p_{k,j}(t),\boldsymbol{p_{-k,j}}(t))\text{ }=\text{ }\frac{\phi_{k,j}(p_{k,j}(t))(1\!+\!I^r_{k,j}(\boldsymbol{p}_{-k,j}(t)))}{\mathcal{R}_k(t,\hat{I}^f(t))x_j(t)}+\text{ }\psi_{k,j}(Q_{k,j}(t)), \label{inst_global_cost}
    \end{align}
    \noindent {where  $I^r_{k,j}(\boldsymbol{p}_{-k,j}(t))$ denotes the expected amount of overlapping content per unit storage size, $C_{k,j}$, $\boldsymbol{p}_{-k,j}(t)$ is a vector of caching control variable of all the other SBSs except SBS $k$,  
    $\hat{I}^f(t)$ denotes the normalized aggregate interference from other SBSs with respect to the SBS density and the number of antennas, and $\mathcal{R}_k(t)$ is the average downlink rate per unit bandwidth.  The cost increases with the amount of overlapping contents and aggregate interference, which are described in the next subsection.} From the global cost function \eqref{inst_global_cost}, the LRA caching cost is given by:
    \begin{align}
    \mathcal{J}_{k,j} = \mathbb{E} \left[\int_t^T J_{k,j}(p_{k,j}(t),\boldsymbol{p_{-k,j}}(t)) \text{ d}t \right]. \label{LRA cost}
    \end{align}
    
    \subsection{Interactions Through Content Overlap and Interference}
    
    The caching strategy of an SBS inherently makes an impact on the caching control strategies of other SBSs. These {\it interactions} can be defined and quantified by the amount of overlapping contents and interference. These represent major bottlenecks for optimizing distributed caching for two reasons: first of all, they undergo changes with respect to the before-mentioned spatio-temporal dynamics, and it is hard to acquire the knowledge of other SBSs's caching strategies directly.
    In this context, our purpose is to estimate these interactions in a distributed fashion without full knowledge of other SBSs' states or actions. 
    
    \textbf{Content Overlap.}\quad
     As shown in Fig. \ref{system_model_CRP}a, in UDNs, there may be overlapping contents downloaded by multiple SBSs located within radius $R_c$ from the randomly selected typical user.
    For example, let us consider that these neighboring SBSs cache the most popular contents with the intention of increasing the local caching gain (i.e., cache hit). Since only one of the SBS candidates is associated with the user to transmit the cached content file, caching the identical content of other SBSs becomes a waste of storage and backhaul usage.
    In this context, overlapping contents increase redundant cost due to inefficient resource utilization \cite{social_cost1}. The amount of overlapping contents is determined by other SBSs' caching strategies. 
    We define the content overlap function $I^r_{k,j}(\boldsymbol{p}_{-k,j}(t))$ as the expected amount of overlapping content per unit storage size $C_{k,j}$, which is given by:
    \begin{align}
    I^r_{k,j}(\boldsymbol{p}_{-k,j}(t))=\frac{1}{C_{k,j}N_{r(j)}}\sum^{|\mathcal{N}|}_{i\neq k} {p}_{i,j}(t), \label{interaction_1}
    \end{align}
    \noindent where 
    $N_{r(j)}$ denotes the number of contents whose request probability is asymptotically equal to $x_j$. It can be defined as  cardinality of the following set: $\{ m | m \in {M} \text{ s.t. } |x_m-x_j| \leq \epsilon\}$. When the value of $\epsilon$ is sufficiently small, $N_{r(j)}$ becomes the number of contents whose request probability is equal to that of content $j$.
    If there is a large number of contents with equal request probabilities, a given content is randomly selected and cached. Hence, the occurrence probability of content overlap decreases with a higher diversity of content caching. 
    
    
    
    \textbf{Inter-SBS Interference.}\quad
    In a UDN, user location determines the  location of interferers, or the density of the user determines the density of interfering SBSs. It is because there are SBSs that have no users in their own coverage and become dormant without imposing interference to their neighboring SBSs. 
    These spatial dynamics of interference in UDN is a bottleneck for optimizing distributed caching such that an SBS in a high interference environment cannot deliver the cached content to its own users. 
    To incorporate this spatial interaction, following the interference analysis in UDNs \cite{JHP1}, interference normalized by SBS density and the number of antennas is given by: 
    \begin{align}
    \hat{I}^f\!(t)\!=\!(\lambda_u\pi R)^2 N_a^{\!-\frac{1}{2}}\lambda_b^{-\frac{\alpha}{2}}\! \left(\!1\!+\! \frac{1-R^{2-\alpha}}{\alpha-2}\! \right)\!{P}\mathsf{E}_g [|g(t)|^2], \label{interaction_2}
    \end{align}
    where $\hat{I}^f(t)$ denotes the normalized interference with respect to SBS density and the number of antennas.
    It gives us an average downlink rate per unit bandwidth $\mathcal{R}_k(t)$ and its upper bound in UDN as follows:
    \begin{eqnarray}
    \mathcal{R}_k(t)&=&\mathsf{E}_{S,I^f}\left[\log(1+\mathsf{SINR}(t))\right] \\ &\leq& \mathsf{E}_{S}\log\left(1+\frac{S_k(t)}{\frac{\sigma^2}{ N_a \lambda_b^{\alpha /2}} + \mathsf{E}_{I^f}[{\hat{I}}^f(t)]}\right), \label{ER_1}
    \end{eqnarray}
    \noindent where $\sigma^2$ is the noise power. Note that inequation \eqref{ER_1}  shows the effect of interference on the upper bound of an average SE. It is because we consider that only the SBSs within the pre-defined reception ball cause interference to a typical user. Hence, the equality in \eqref{ER_1} holds, when the size of reception ball $R$ goes to infinity, including all the SBSs in the networks as interferers.

    \subsection{Stochastic Differential Game for  Edge Caching}
    
    As the SBSs' states and content popularity evolves according to the dynamics \eqref{SDE_x} and \eqref{SDE_q}, an individual SBS's caching strategy must adapt accordingly. Hence, minimizing the LRA cost under the spatio-temporal dynamics can be modeled as a dynamic stochastic differential game (SDG), where the goal of each SBS $k$ is to determine its own caching amount ${p}_{k,j}^*(t)$ for content $j$  to minimize the LRA cost  $\mathcal{J}_{k,j}(t)$ \eqref{LRA cost}:
    \begin{align}
    \hspace{-1.3cm}\textbf{(P1)} \quad\quad&v_{k,j}(t)=\mathop{\text{inf}}\limits_{p_{k,j}(t)}\text{ } \mathcal{J}_{k,j}(t) \\
    \text{subject to }\quad\quad&\text{d}x_j(t)= r(\mu-x_j(t))\text{d}t+\eta \text{d}W_j(t), \quad\label{const_1} \\ 
    &\text{d}Q_{k,j}(t)=(e_{k,j}-L_j p_{k,j}(t))\text{d}t.   \label{const_2}
    \end{align} 
    In the problem \textbf{P1}, the state of SBS $k$ and content $j$ at time $t$ is defined as 
    $\boldsymbol{s}_{k,j}(t)=\{x_j(t),\mathcal{R}_k(t),Q_{k,j}(t)\}$, $\forall k \in \mathcal{N}, \forall j \in \mathcal{M}$. The stochastic differential game (SDG) for edge caching is defined by
    $(\mathcal{N},\mathcal{S}_{k,j}, \mathcal{A}_{k,j}, \mathcal{J}_{k,j}    )$ where $\mathcal{S}_{k,j}$ is the state space of SBS $k$ and content $j$, $\mathcal{A}_k$ is the set of all caching controls $\{p_{k,j}(t), 0 \leq t \leq T \}$ admissible for
    the state dynamics.

    To solve the problem \textbf{P1}, the long-term average of content request probability $\mu$ is necessary for the dynamics of content request probability \eqref{const_1}. 
    To determine the value of $\mu$, the mean value $m_k(t)$ of the cardinality of the set $U_k^r(t)$ needs to be obtained. 
    Although the period $\{0\leq t\leq T\}$ is not infinite, we assume that the inter-arrival time of the content request is sufficiently smaller than $T$ and that numerous content requests arrive during that period. Then, the long-term average of content request probability $\mu$ becomes an asymptotic mean value $(t\!\rightarrow\! \infty)$.
    Noting that $\sum_j n_{k,j}(t) = N_k(t)$, the mean value of $m_k(t)$ is asymptotically given by \cite{CRP_2} as follows:
    \begin{align}
    \left<|U^r_k(t)|\right>  \simeq \left\{ \begin{array}{ll}
     \frac{\Gamma(\theta+1)}{\alpha\Gamma(\theta+\alpha)}N_k(t)^{\alpha} & \textrm{for $\alpha>0$}\\
    \theta \log(N_k(t)+\theta) & \textrm{for $\alpha=0$}\end{array} \right.  \label{avg_Nk}
    \end{align}
    where the expression $\left<.\right>$ is the average value, and $\Gamma(.)$ is the Gamma function. 

    The problem \textbf{P1} can be solved by using a backward induction method where the minimized LRA cost $v_{k,j}(t)$ is determined in advance through solving the following  $N$ coupled HJB equations. 
    \begin{align}
    0&\!= \partial_t v_{k,j}(t)\! +\!\mathop{\text{inf}}\limits_{p_{k,j}(t)} \bigg{[}J_{k,j}(p_{k,j}(t),\boldsymbol{p_{-k,j}}(t))\!+\! \frac{\eta^2}{2}\partial_{xx}^2 v_{k,j}(t)\nonumber \\
     &+\! \underbrace{(e_{k,j} -L_j p_{k,j}(t))}_{(A)} \partial_{Q_k}v_{k,j}(t)+\underbrace{r(\mu-x_j(t))}_{(B)}\partial_x v_{k,j}(t)\bigg{]} \label{hjb_sdg}
    \end{align}
    
    {The HJB equations \eqref{hjb_sdg} for $ k=1,...,N$ have a unique joint solution if the drift functions defining temporal dynamics (A) and (B) and the cost function \eqref{inst_global_cost} are smooth \cite{exist_HJBsol1}. Since the smoothness of them is satisfied, we can assure that a unique solution of equation \eqref{hjb_sdg} exists. The optimal joint solution of HJB equations achieves Nash equilibrium (NE) as the problem \textbf{P1} is a non-cooperative game wherein players do not share their state or strategy  \cite{exist_HJBsol1,EV_JSAC}. 
    The unique minimized cost $v^*_{k,j}(t)$ of the problem \textbf{P1} and its corresponding NE can be defined as follows:
    }
    
    \vskip 10pt
    \noindent {\bf Definition 1}: The set of SBSs' caching strategies $\mathbf{p}^*=\{p_{1,j}^*(t),...,p_{N,j}^*(t) \}$, where $p_{k,j}^*(t) \in  \mathcal{A}_{k,j}$ for all $k \in \mathcal{N}$, is a Nash equilibrium, if for all SBS $k$ and for all admissible caching strategy set $\{p_{1,j}(t),...,p_{N,j}(t) \} $, where $p_{k,j}(t) \in  \mathcal{A}_{k,j}$ for all $k \in \mathcal{N}$, it is satisfied that 
    \begin{align}
    \mathcal{J}_{k,j}(p_{k,j}^*(t), \boldsymbol{p}_{-k,j}^*(t))\leq \mathcal{J}_{k,j}(p_{k,j}(t), \boldsymbol{p}_{-k,j}^*(t)), \label{Def_NE_SDG}
    \end{align}
    under the temporal dynamics \eqref{const_1} and \eqref{const_2} for common initial states $x_j(0)$ and $Q_{k,j}(0)$.
    
    \vskip 10pt
    
    {Unfortunately, this approach is accompanied with high computational complexity in achieving the NE \eqref{Def_NE_SDG}, when $N$ is larger than two because an individual SBS should take into account other SBSs' caching strategies $\boldsymbol{p_{-k,j}}(t)$ 
    to solve the inter-weaved system of $N$ HJB equation \eqref{hjb_sdg}.  Furthermore, it requires collecting the control information of all other SBSs including their own states, which brings about a huge amount of information exchange among SBSs. This is not feasible and impractical for UDNs. For a sufficiently large number of SBSs, this problem can be transformed  to a mean-field game (MFG), which can achieve the $\epsilon$-Nash equilibrium~\cite{MFG_application}.}
    
    \subsection{Mean-Field Game for Edge Caching} \label{Mean-field Game for Caching}

    To reduce the aforementioned complexity in solving the SDG \textbf{P1}, the following features are utilized.
    When the number of SBSs becomes large, the influence of every individual SBS can be modeled with the effect of the collective (aggregate) behavior of the SBSs. 
    {MFG theory enables us to transform these multiple interactions into a single aggregate interaction, called MF interaction, via MF approximation. 
    According to \cite{MF_ref2}, this approximation holds under the following conditions: 
    (i) a large number of players, (ii) the exchangeability of players under the caching control strategy, and (iii) finite MF interaction. If these conditions are satisfied, the MF approximation can provide the optimal solution which the original SDG achieves.}
    
    The first condition (i) corresponds to the definition of UDNs. For condition (ii),
    players (i.e., SBSs) in the formulated SDG are said to be exchangeable or indistinguishable under the control ${p}_{k,j}(t)$ and the states of players and contents 
    if the player's control is invariant by their indices and decided by only their own states.
    In other words, permuting players' indices cannot change their own control strategies. Under this exchangeability, it is sufficient to investigate and re-formulate the problem for a generic SBS by dropping its index~$k$. 
    
    The MF interactions \eqref{interaction_1} and \eqref{interaction_2} should asymptotically converge to a finite value under the above conditions.
    The content overlap \eqref{interaction_1} in MF regime, called MF overlap, goes to zero when the number of contents per SBS is extremely large, i.e. $M \gg N $. Such a condition implies that the cardinality of the set consisting of asymptotically equal content popularity goes to infinity. In other words, $N_{r(j)}$ goes to infinity yielding that the expected amount of overlapping content per unit storage size $I^r_{k,j}(\boldsymbol{p}_{-k,j}(t))$ becomes zero. 
    In terms of interference, the MF interference converges  as
    the ratio of SBS density to user density goes to infinity, i.e.  $N_a\lambda_b^{\alpha}/(\lambda_u R)^4 \rightarrow \infty$ \cite{JHP1}. This condition corresponds to the notion of UDN  \cite{interf_udn} or massive MIMO ($N_a \rightarrow \infty$). Thus, the MF approximation can be utilized as the conditions inherently hold for UDCNs .
    
    To approximate the interactions from other SBSs, we need a state distribution of SBSs and contents at time $t$, called MF distribution $m_t(x(t),Q(t))$.
    The MF distribution is derived from the following empirical distribution.
    
    \begin{align}
     M_t^{(N\times M)}(x(t),Q(t))= \frac{1}{NM}\sum_{j=1}^{M}\sum_{k=1}^{N} \delta_{\{x_j(t),Q_k(t)\}}
    \end{align}
    When the number of SBSs increases, the empirical distribution $M_t^{(N \times M)}(x_j(t),Q(t))$ converges to $m_t(x_j(t),Q(t))$, which is the density of contents and SBSs in state $(x_j(t),Q(t))$. Note that we omit the SE $\mathcal{R}(t)$ from the density measure to only consider temporally correlated state without loss of generality.
    
    To this end, we derive a Fokker-Planck-Kolmogorov (FPK) equation \cite{MFG_application} that is a partial differential equation capturing the time evolution of the MF distribution $m_t(x_j(t),Q(t))$ under dynamics of the popularity $x_j(t)$ and the available storage size $Q(t)$. 
    The FPK equation for $m_t(x_j(t),Q(t))$ subject to the temporal dynamics  \eqref{SDE_x} and  \eqref{SDE_q} are given as follows:
    \begin{align}
    &0= \partial_t m_t(x_j(t),Q(t)) +r(\mu_j-x_j(t))\partial_x m_t(x_j(t),Q(t)) \nonumber \\&+ (e_j\!-\!L_j p_{j}(t)) \partial_{Q}m_t(x_j(t),Q(t)) 
     \!- \!\frac{\eta^2}{2}\partial_{xx}^2 m_t(x_j(t),Q(t)).  \label{fpk_1}
    \end{align}
    \noindent Let us denote the solution of the FPK equation \eqref{fpk_1} as $m_t^*(x_j(t),Q(t))$. 
    Exchangeability and existence of the MF distribution allow us to approximate the interaction $I^r_{k,j}(\boldsymbol{p}_{-k,j}(t))$ as a function of $m_t^*(x_j(t),Q(t))$ as follows:
    \begin{align}
    I^r_{j}(t,m_t^*(x_j(t),Q(t)\!)\!)\!=\!\!\int_{\!Q}\!\int_{\!x} \frac{m_t^*(x_j,Q){p}_{j}(t,\!x(t),\!Q(t))}{C_{k,j}N_{r(j)}} \text{d}x\text{d}Q. \label{MF_interaction_apprx}
    \end{align}
    {\noindent This interaction from \eqref{MF_interaction_apprx} can be estimated without observing other SBSs' caching strategies.
    Thus, it is not necessary for an SBS to have full knowledge of the states or the caching control policies of other SBSs. An SBS needs to solve only a pair of equations, namely the FPK equation \eqref{fpk_1} and the following modified HJB one obtained by applying the MF approximation \eqref{MF_interaction_apprx} to \eqref{hjb_sdg}: } 
    \begin{align}
    0&\!= \partial_t v_{j}(t)\! +\!\!\mathop{\text{inf}}\limits_{p_{j}(t)} \!\bigg{[}\!J_{j}(p_{j}(t),I_j(t,m_t^*(x_j(t),Q(t)\!)\!)\!+\! \frac{\eta^2}{2}\partial_{\small{xx}}^2 v_{j}(t)\nonumber \\
     &+ (e_j -L_j p_{j}(t)) \partial_{Q}v_{j}(t)+r(\mu-x_j(t))\partial_x v_{j}(t)\bigg{]}. \label{hjb_mfg}
    \end{align}
    
    { FPK equation \eqref{fpk_1} and HJB equation \eqref{hjb_mfg} are intertwined with each other for the MF distribution and the optimal caching amount, which depends on the optimal trajectory of the LRA cost $v_j^*(t)$.
    The optimal LRA cost  $v^*_{j}(t)$ is found by applying backward induction to the single HJB equation \eqref{hjb_mfg}. Also, its corresponding MF distribution (state distribution) $m_t^*(x_j(t),Q(t))$ is obtained by forward solving the FPK equation \eqref{fpk_1}. These solutions of HJB and FPK equations $[m_t^*(x_j(t),Q(t)), v_j^*(t)]$ define the mean-field equilibrium (MFE)}, defined as follows:
    
    \vskip 10pt
    \noindent {\bf Definition 2}: The generic caching strategies $p_{j}^*(t)$ achieves an  MFE if for all admissible caching strategy set $\{p_{1,j}(t),...,p_{N,j}(t) \} $ where $p_{k,j}(t) \in  \mathcal{A}_{k,j}$ for all $k \in \mathcal{N}$ it is satisfied that 
    \begin{align}
    \mathcal{J}_{j}(p_{j}^*(t), m^*_t(x_j(t),Q(t))\leq \mathcal{J}_{j}(p_{j}(t), m^*_t(x_j(t),Q(t)), \label{Def_NE}
    \end{align}
    under the temporal dynamics \eqref{const_1} and \eqref{const_2} for an initial MF distribution $m_0$.
    The MFE corresponds to the $\epsilon$-Nash equilibrium:
    \begin{align}
    \mathcal{J}_{k,j}(p_{k,j}^*(t), \boldsymbol{p}_{-k,j}^*(t))\leq \mathcal{J}_{j}(p_{j}^*(t), m^*_t(x_j(t),Q(t)) -\epsilon, 
    \end{align}
    where $\epsilon$ asymptotically becomes to zero for a sufficiently large number of SBSs.
    
    Let us define $p_j^*(t)$ as an optimal caching control strategy which achieves the MFE yielded by the optimal caching cost trajectory $v_j^*(t)$ and MF distribution $m_t^*(x_j(t),Q(t))$.
    The solution $p_j^*(t)$ is given by the following Proposition.
    %

    \vskip 10pt
     \noindent {\bf Proposition 1.} {\it The optimal caching amount is given by: }
    \begin{eqnarray}
    p_{j}^*(t)=\frac{1}{L_j}\left[B_{j}(t)- \frac{1+I^r_j(t,m^*_t(x_j(t),Q(t)))}{\mathcal{R}(t,I^f(t)) x_j(t)\partial{\scriptscriptstyle 
     {Q}}{v^*_{j}}}    \right]^+,  \label{Propo1}
     \end{eqnarray}
     {\it where $m_t^*(x(t),Q(t))$ and $v_j^*(t)$ are the  solutions of  \eqref{fpk_1} and \eqref{hjb_mfg}, respectively. }  
     \vskip 10pt
    \noindent {\it Proof}: The optimal control control of the differential game with HJB equations  is the argument of the infimum term \eqref{hjb_mfg}~\cite{exist_HJBsol1}.
    \begin{align}
    p_j^*(t)&\!=\!\arginf\limits_{p_{j}(t)} \bigg{[}J_{j}(p_{j}(t),\!I_j(t,m_t^*(x_j(t),Q(t)))\!+\! \frac{\eta^2}{2}\partial_{xx}^2 v_{j}(t)\nonumber \\
     &+\! (e_j -L_j p_{j}(t)) \partial_{Q}v_{j}(t)+r(\mu-x_j(t))\partial_x v_{j}(t)\bigg{]} \label{infimum}
    \end{align}
    The infimum term \eqref{infimum} is a convex function of $p_j(t)$ for all time $t$, since its first and second-order derivative are lower than zero. Hence, we can apply Karush-Khun-Tucker (KKT) conditions and get a sufficient condition for the unique optimal control $p_j^*(t)$ by finding a critical point given by:
    \begin{align}
    \frac{\partial}{\partial p_j(t)} \big[J_{j}(p_{j}(t),I_j(t,m_t^*(x_j(t),Q(t))) \nonumber\\
    + (e_j -L_j p_{j}(t)) \partial_{Q}v_{j}(t) \big]= 0.\label{pppp}
    \end{align}
    Due to the convexity, the solution of equation \eqref{pppp} is the unique optimal solution described as follows:
    \begin{align}
    p_{j}^*(t)=\frac{1}{L_j}\left[B_{j}(t)- \frac{1+I^r_j(t,m^*_t(x_j(t),Q(t)))}{\mathcal{R}(t,I^f(t)) x_j(t)\partial{\scriptscriptstyle 
     {Q}}{v^*_{j}}}    \right]^+.
    \end{align}
    Remark that $p^*_j(t)$ is a function of $m^*_t(x_j(t),Q(t))$ and $v^*_{j}$, which are solutions of the equations \eqref{fpk_1} and \eqref{hjb_mfg}, respectively.
    The expression of $p_j^*(t)$ \eqref{pppp} provides the final versions of the HJB and FPK equations as follows:
    
    {
    \begin{align}
    0&= \partial_t v_{j}(t) -\frac{\log \left(B_{j}(t)-\left[B_{j}(t)- \frac{1+I^r_j(t,m_t^*(x_j(t),\!Q(t)))}{\mathcal{R}(t,I^f(t))x_j(t)\partial{\scriptscriptstyle{Q}}{v_{j}}}    \right]^{\scriptscriptstyle{+}}\right)}{\mathcal{R}(t,I^f(t))x_j(t)} \nonumber \\
     &\!\times \!(1\!+\!I^r_j(t,\!m_t^*(x_j(t),\!Q(t)\!)\!)\!)+\!\frac{\alpha(C\!-\!Q(t))}{C}\!+\!r(\mu_j\!-\!x_j(t)\!)\partial_x v_{j}(t)\nonumber \\
     &\!+\! \left(\!e_j \!-\!\left[\!B_{j}(t)\!- \!\frac{\!1\!+\!I^r_j(t,m_t^*(x,\!Q))}{\mathcal{R}(t,I^f(t))x_j(t)\partial{\scriptscriptstyle{Q}}{v_{j}}}\!\right]^{\!\scriptscriptstyle{+}}\right) \!\partial_{Q}v_{j}(t)\!+\! \frac{\eta^2}{2}\partial_{xx}^2 v_{j}(t), \nonumber\\
     \nonumber\\
    0&= \partial_t m_t(x_j(t),Q(t)) +r(\mu_j-x_j(t))\partial_x m_t(x_j(t),Q(t)) 
    \nonumber\\&- \frac{\eta^2}{2}\partial_{xx}^2 m_t(x_j(t),Q(t))
    \nonumber \\&+\! \left(\!e_j \!-\!\left[\!B_{j}(t)\!- \!\frac{\!1\!+\!I^r_j(t,m_t(x_j(t),Q(t)))}{\mathcal{R}(t,I^f(t))x_j(t)\partial{\scriptscriptstyle{Q}}{v_{j}^*}}\!\right]^{\!\scriptscriptstyle{+}}\right) \!\partial_{Q}m_t(x_j(t),\!Q(t)). \nonumber
    \end{align}}
    
    \noindent From these equations, we can find the values of
    $v_j^*(t)$ and $m_t^*(x(t),Q(t))$. Note that the smoothness of the drift functions and in the dynamic equation and the cost function \eqref{inst_global_cost} assures the uniqueness of the solution \cite{exist_HJBsol1}. \hfill$\blacksquare$ 
    \vskip 10pt
    
    \begin{figure}
    \centering
    \includegraphics[width=\textwidth]{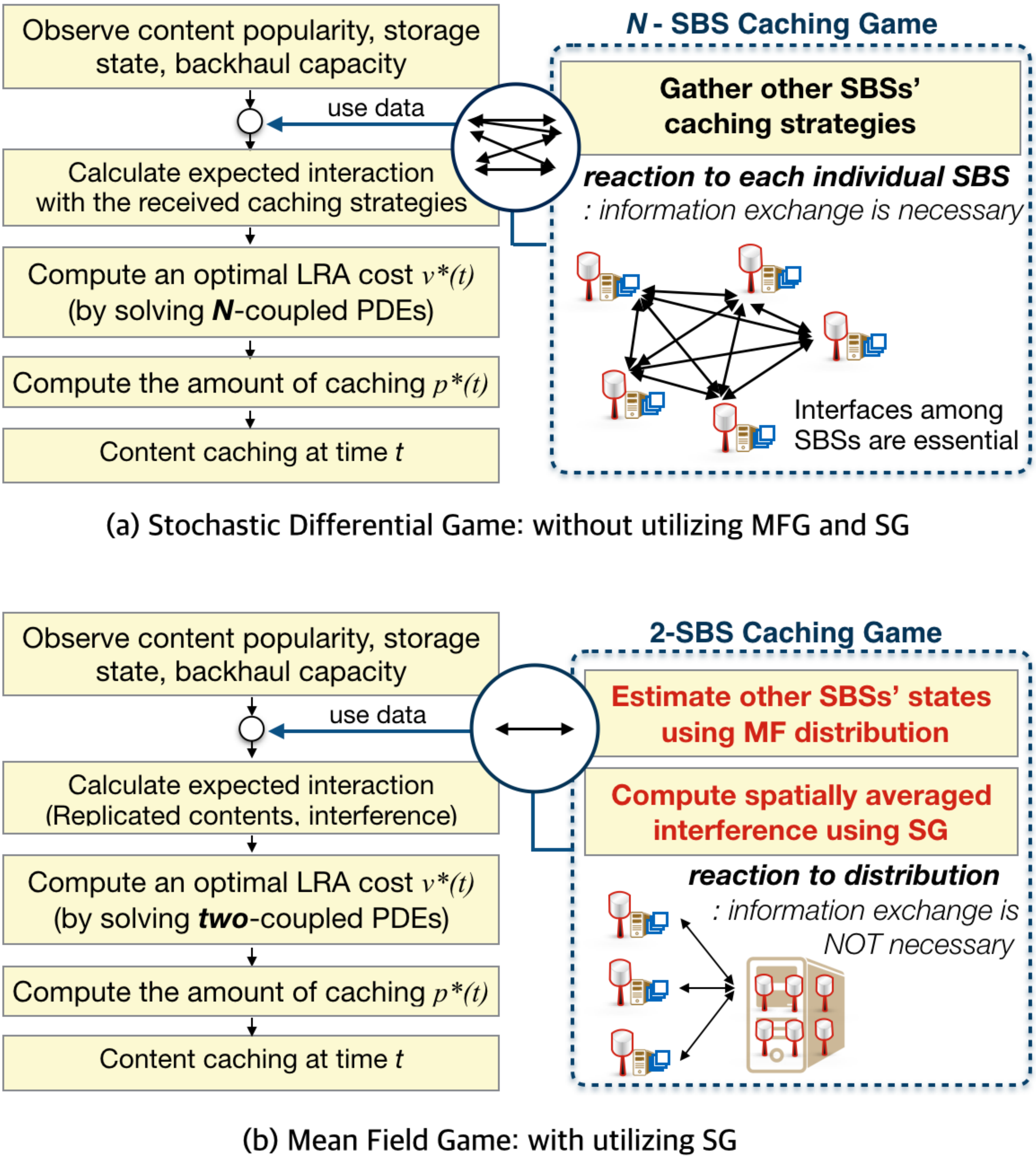}   
    \caption{\small{Ultra-dense edge caching flow charts according to the approaches of SDG and MFG, respectively. (a) In the framework of SDG, we solve the game of $N$ SBSs (players)  interacting with each individual SBS. (b) By incorporating  MFG theory and SG into the framework, we can estimate the collective interaction of other SBSs. This relaxes the $N$-SBS caching game to a two-SBS caching game. }  }\label{diagram} 
    \end{figure}

    Proposition 1 provides the optimal caching amount of $p_{j}^*(t)$ is in a water-filling fashion of which water level is determined by the backhaul capacity $B_j(t)$. Noting that the average rate per unit bandwidth $\mathcal{R}(t)$ increases with the number of antenne $N_a$ and SBS density $\lambda_b$, SBSs cache more contents from the server when they can deliver content to users with high wireless capacity. 
    Also, SBSs diminish the caching amount of content $j$, when the estimated amount of content overlap $I^r_j(t,m^*_t(x_j(t),Q(t)))$ is large.

    Remark that the existence and uniqueness of the optimal caching control strategy are guaranteed. The optimal caching algorithm converges to a unique MFE, when the initial conditions $m_0$, $x_j(0)$, and $Q(0)$ are given.
     The specific procedure of this MF caching algorithm is described in the following Algorithm~1.

     \begin{algorithm}
        \centering
        \caption{Mean-Field Caching Control}\label{euclid}
        \begin{algorithmic}[1]
        \REQUIRE $x_j(t)$, $m_0$, $B(t)$ and $Q(0)$
        \STATE  Find the optimal trajectory of caching cost and state distribution $[v_j^*(t),m_t^*(x_j(t),Q(t))]$ by solving HJB \eqref{hjb_mfg} and FPK \eqref{fpk_1} equations \vskip 3pt
        \STATE Calculate  $I_j^r(t,m_t^*(x_j(t),\!Q(t)))$, $I^f(t)$ and $\partial{\scriptscriptstyle{Q}}{v^*_{j}}$
         \vskip 3pt
        \STATE Compute the instantaneous caching amount $p_{j}^*(t)$\\ :$\quad p_{j}^*(t)=\frac{1}{L_j}\left[B_{j}(t)- \frac{1+I^r_j(t,m^*_t(x_j(t),Q(t)))}{\mathcal{R}(t,I^f(t)) x_j(t)\partial{\scriptscriptstyle 
         {Q}}{v^*_{j}}}    \right]^+$
         \vskip 3pt
        \STATE Get values of $[x_j(t),Q(t)]$  according to the dynamics 
        \STATE Go line 2
        \end{algorithmic}
        \end{algorithm}

        \begin{figure}\centering
            \includegraphics[width=.8\textwidth]{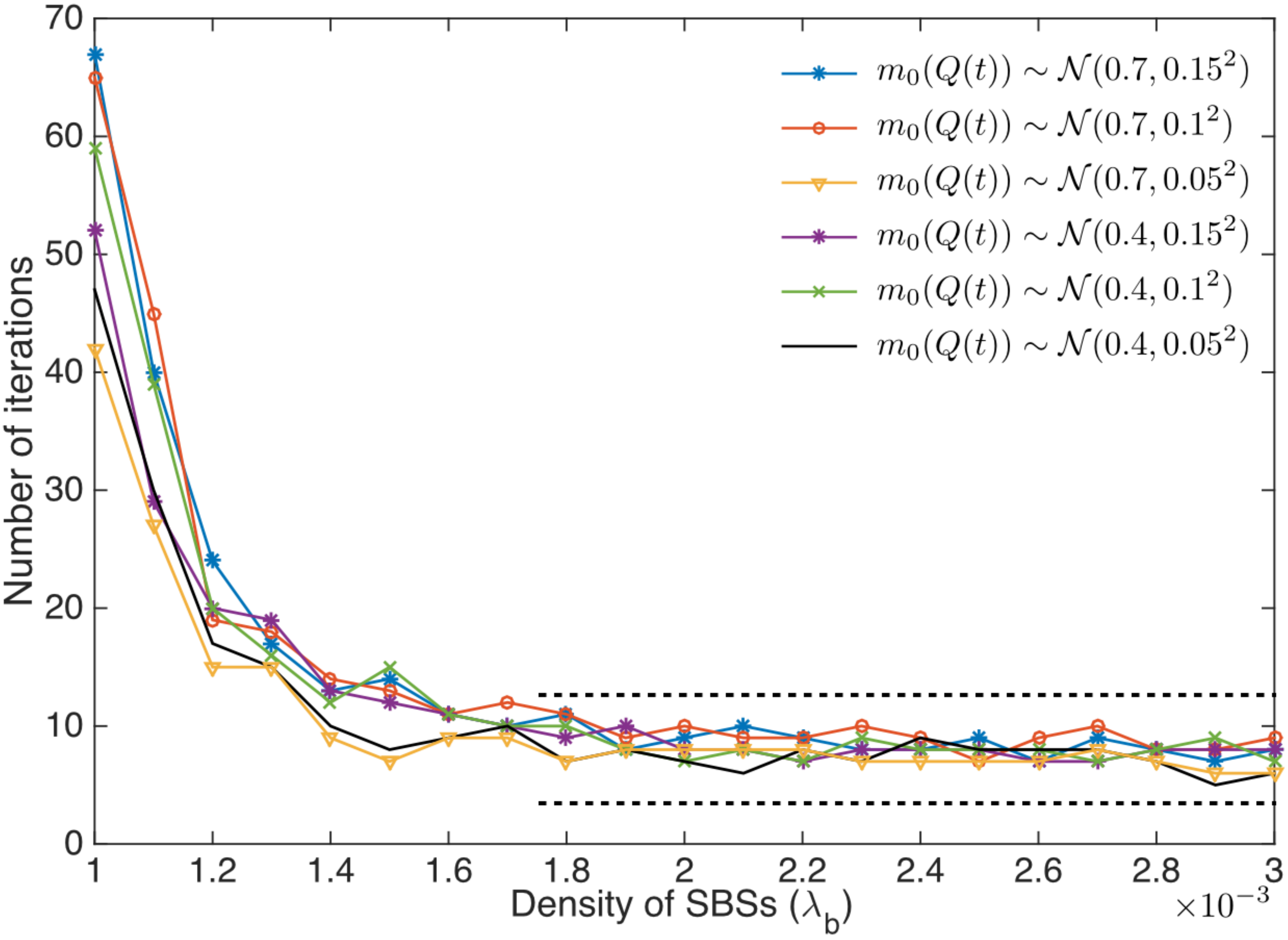}   
            \caption{\small{The number of iterations required to solve the coupled HJB and FPK equations for different densities of SBSs.}  }\label{complexity_it}
            \end{figure}
            
            {The respective processes of solving \textbf{P1} in ways of SDG and MFG are depicted in Fig. \ref{diagram}.  
            Remark that the solution of the MFG becomes equivalent to that of the $N$-player SDG \textbf{P1} as $N$ increases.  
            The complexity of the proposed method is much lower compared to solving the original $N$-player SDG  \textbf{P1}. The number of PDEs to solve for one content is reduced to two from the number of SBSs $N$. Thus, the complexity is consistent even though the number of players $N$ becomes large. }
            {This feature is verified via simulations as shown in Fig. \ref{complexity_it}, which represents the number of iterations required to solve the HJB-FPK equations \eqref{hjb_mfg} and \eqref{fpk_1} as a function for different SBS densities $\lambda_b$. 
            Here, it is observed that the caching problem \textbf{P1} is numerically solved by within a few iterations for highly dense networks. It means that the computational complexity remains consistent regardless of the SBS density $\lambda_b$, or the number of players $N$. Fig. \ref{complexity_it} also shows that this consistency holds for different initial storage state distribution of SBSs. The number of iterations to reach the optimal caching strategy is bounded within tens of iterations even for low SBS density. The proposed algorithm provides the solution faster for more densified networks.}

            \section{Numerical Results}\label{C_numerical}

            Numerical results are provided for evaluating the proposed algorithm under spatio-temporal content popularity and network dynamics illustrated in Fig.~\ref{system_model_CRP}. 
            Let us assume that the initial distribution of the SBSs $m_0$ is given as normal distribution and that the storage size $Q(t)$ belongs to a set $[0,1]$ for all time $t$. Considering Rayleigh fading with mean one, the parameters are configured as shown in Table I.
            {To solve the coupled PDEs (the first step of the Algorithm 1) using a finite element method, we used the MATLAB PDE solver.}

            \begin{table} [h]
            \centering \caption{Key simulation parameters}\small
            \small\begin{tabular}{|l||l|}
              \hline
              \footnotesize Parameter & \footnotesize Value  \\
              \hline
            \footnotesize SBS density $\lambda_b$& \footnotesize 0.005, 0.02, 0.035, 0.05 (SBSs/m$^2$)
             \\
            \footnotesize  User density $\lambda_u$  &\footnotesize $10^{-4}$, $2.5 \times 10^{-4}$ (users/m$^2$)
               \\
            \footnotesize   Transmit power $P$ &\footnotesize 23 dBm
             \\
             \footnotesize  Noise floor &\footnotesize -70 dBm
             \\
            \footnotesize   Number of contents    &\footnotesize 20
             \\
               CRP parameters  $\theta, \nu$  &\footnotesize $\theta=1,\nu=0.5$
               \\
            \footnotesize  Reception ball radius $R$ &\footnotesize $10/\sqrt{\pi} $ km
             \\
            \footnotesize  Network size &\footnotesize 20  km $\times$ 20 km
             \\
            \footnotesize   File discarding rate $e_j$&\footnotesize 0.1
             \\
              \hline
            \end{tabular}\label{table}
            \end{table}
            
            \subsection {Mean-field equilibrium achieved by the proposed MF caching algorithm}

            \begin{figure} 
            \centering
            \includegraphics[angle=0, width=7cm]{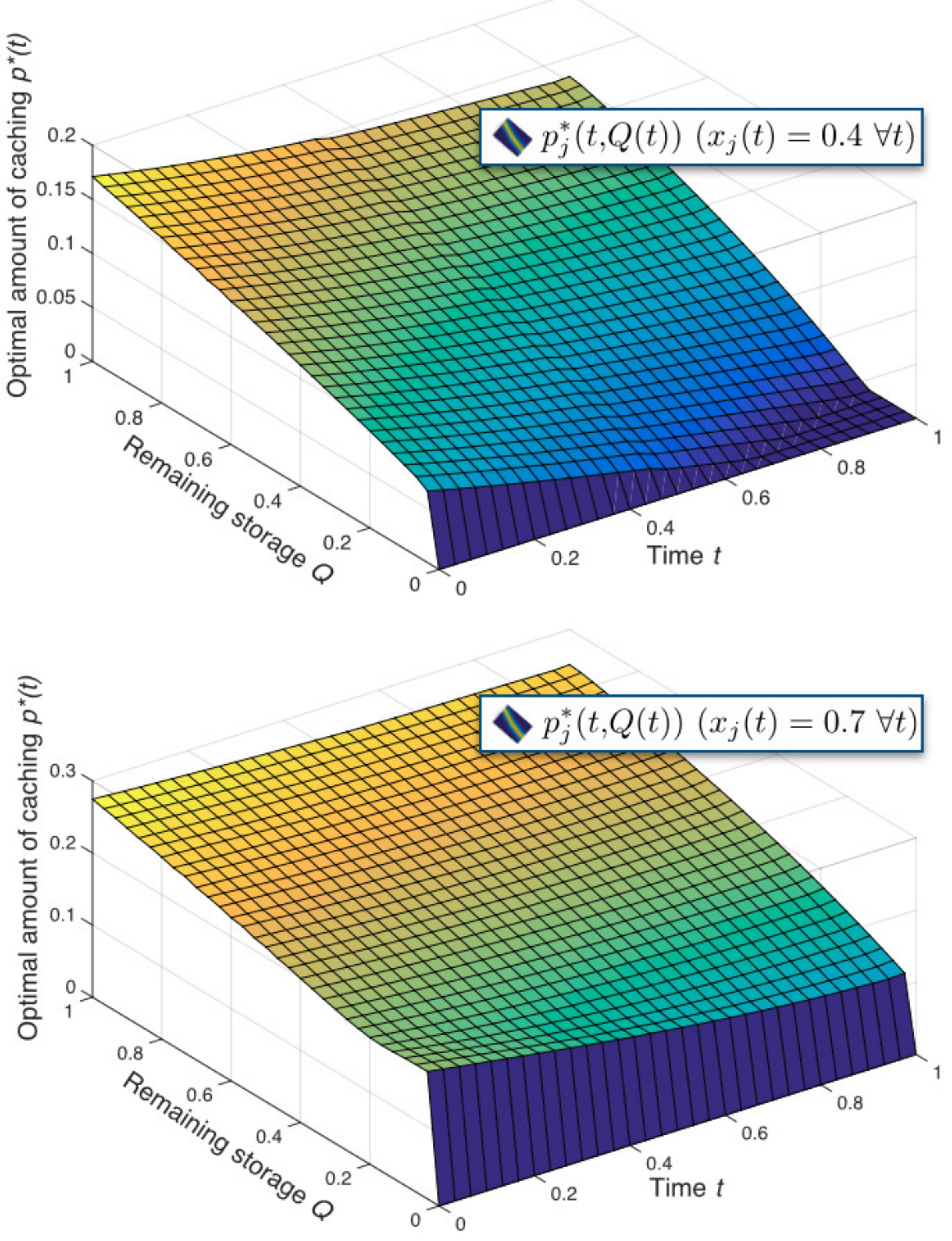}   
            \caption{\small{The optimal caching amount  $p^*(t)$ at the MF equilibrium under two different content popularities 0.4 and 0.7, assuming that the content popularity is static. The initial MF distribution $m_0(Q(0))$ is given as $\mathcal{N}(0.7,0.05^2)$.   }  }\label{trajectory_control}
            \end{figure}

            To demonstrate that the proposed MF caching algorithm achieves the MFE, it is assumed that SBSs have full knowledge of contents request probability, which implies perfect popularity information is available at SBSs.
            The trajectory of the proposed caching algorithm and MF distribution is numerically analyzed when the content request probability is static. 
            In this case,  the caching control strategies do not depend on the evolution law of the content popularity. 
            Specifically, in HJB \eqref{hjb_mfg}  and FPK \eqref{fpk_1} equations, the derivative terms with respect to  content request probability $x$ become zero.

            
            Fig. \ref{trajectory_control} shows the evolution of the optimal caching amount $p^*(t)$ with respect to the storage state and time. The value of $p^*(t)$ is maintained lower than the content request probability to reduce the content overlap and prevent redundant backhaul and storage usage.

            \begin{figure*}[ht]
            \centering
            \includegraphics[width=\textwidth]{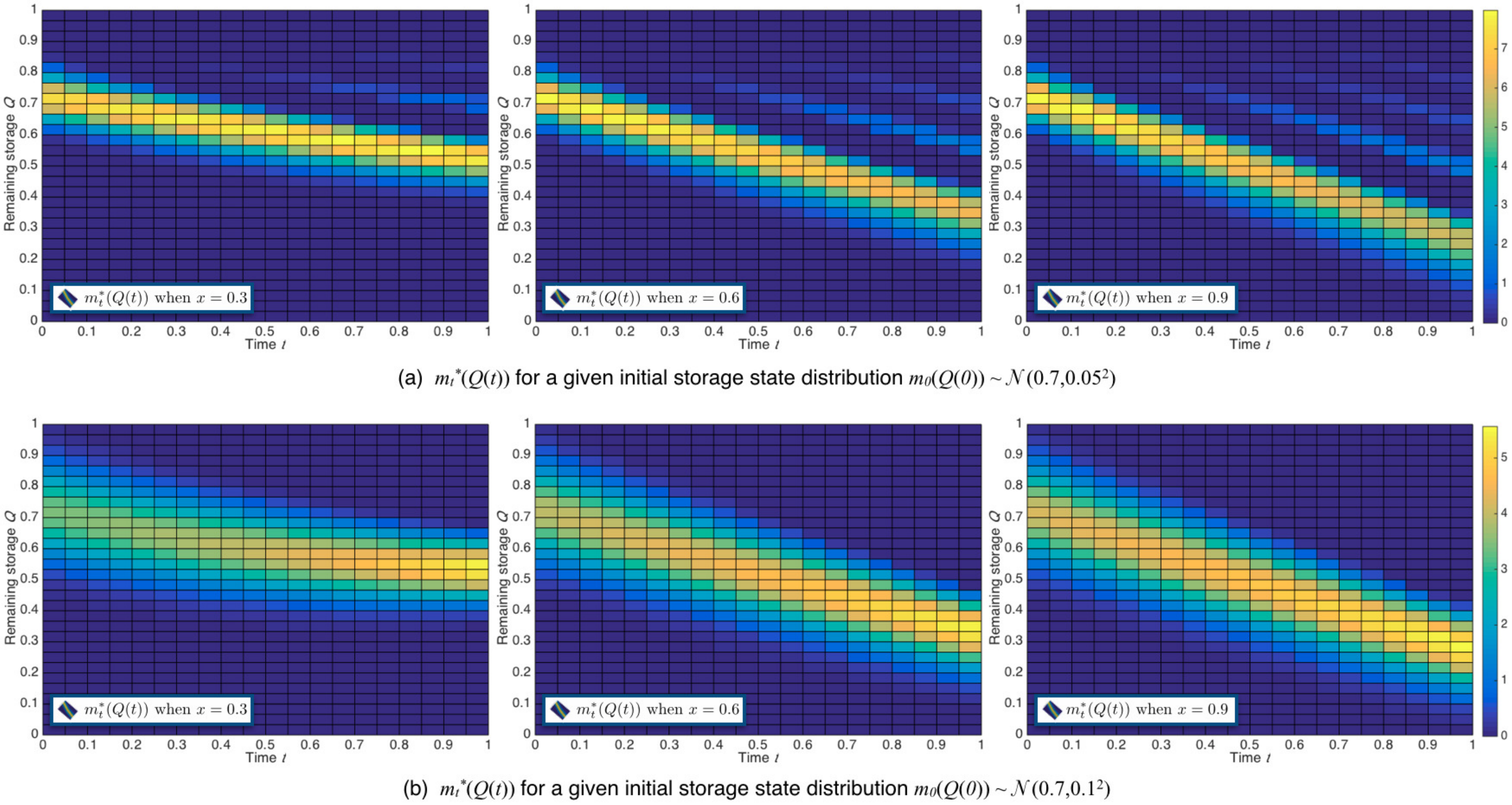} 
            \caption{{\small A heat map illustration of the MF distribution $m_t^*(Q(t))$ that represents the instantaneous density of SBSs having the remaining storage space $Q(t)$  for an arbitrary content during a long-term period $\{\ 0\leq t \leq T\}$, when the proposed MF caching algorithm is applied. 
            A bright-colored point means there are many SBSs with the unoccupied storage size corresponding to the point.
            It shows the temporal evolution of the density of SBSs  with respect to different content popularity $x_j$, and initial distribution $m_0(Q(0))$ $(B(t)=1,N_{r(j)}=20, \lambda_u=0.001,\lambda_b=0.03)$.}}\label{MF_dist_total} 
            \end{figure*}

            Fig. \ref{MF_dist_total}  shows heat-maps representing the instantaneous density of SBSs having the remaining storage size $Q(t)$ in terms of the MF distribution $m_t^*(Q(t))$ for a content during a period $\{0\leq t \leq T\}$, where $T=1$. A bright-colored point means there are many SBSs with the unoccupied storage size corresponding to the point. It is observed that the unoccupied storage space of SBSs does not diverge from each other as the proposed algorithm brings SBSs' state in the MFE.
            At this equilibrium, the amount of cached content file decreases when the content popularity $x$ becomes low. This tendency corresponds to the trajectory of the optimal caching probability in Fig. \ref{trajectory_control}.
            Almost every SBS has cached the content over time, but not used its entire storage. The remaining storage saturates even though the content popularity is equal to $0.9$. 
            This implies that SBSs adjust the caching amount of popular content in consideration of the content overlap expected to possibly increase the cost.  
            
            \subsection {Performance evaluation in terms of long run average caching cost}

            \begin{figure}
            \centering
            \includegraphics[width=.8\textwidth]{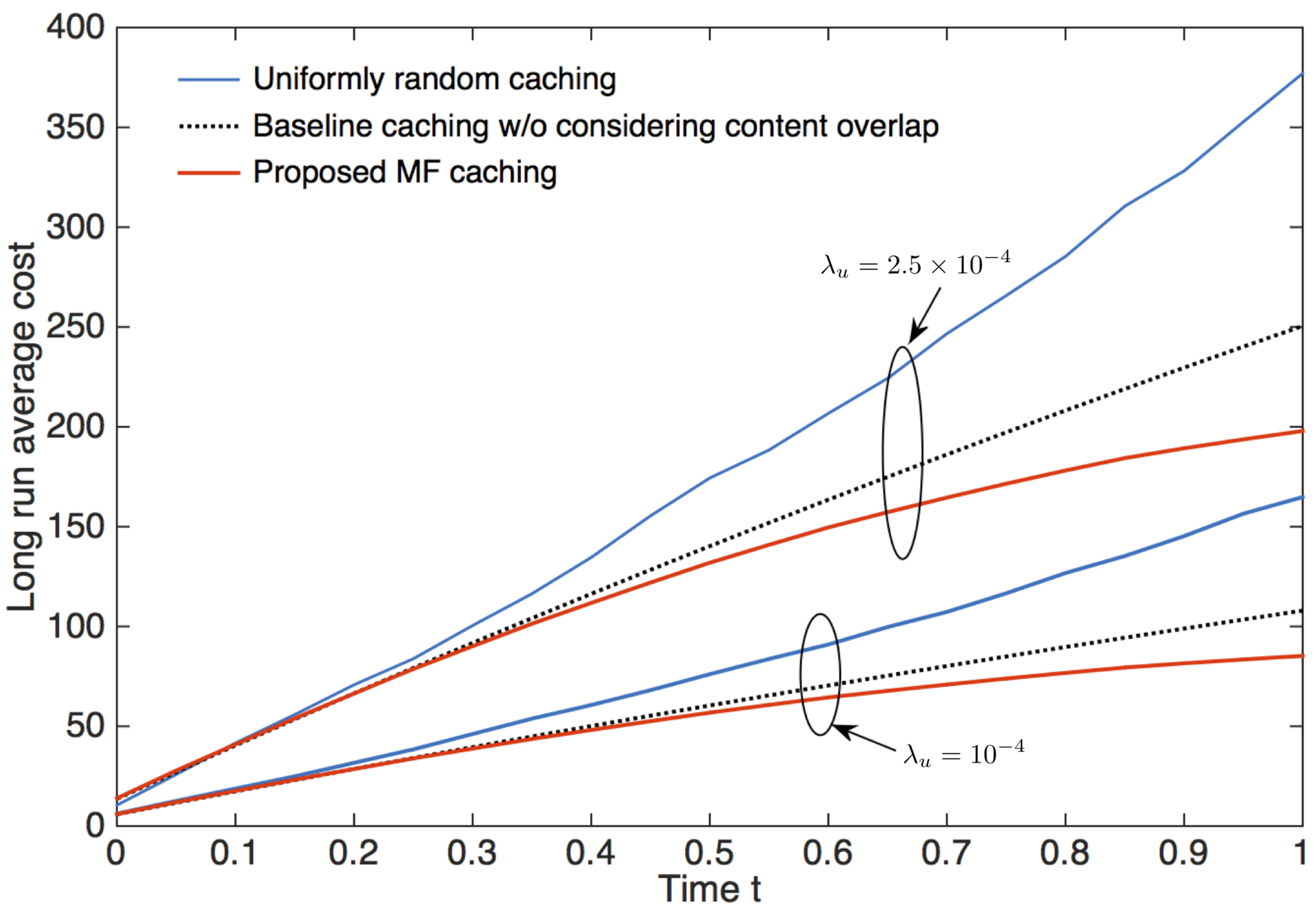}   
            \caption{{\small Long run average costs of the caching strategies with respect to different user density $\lambda_u$. ($Q(0)=0.7, x(0) =0.3, \eta=0.1$).} }\label{LRA_user} 
            \end{figure}
            
            \begin{figure}
            \centering
            \includegraphics[width=.8\textwidth]{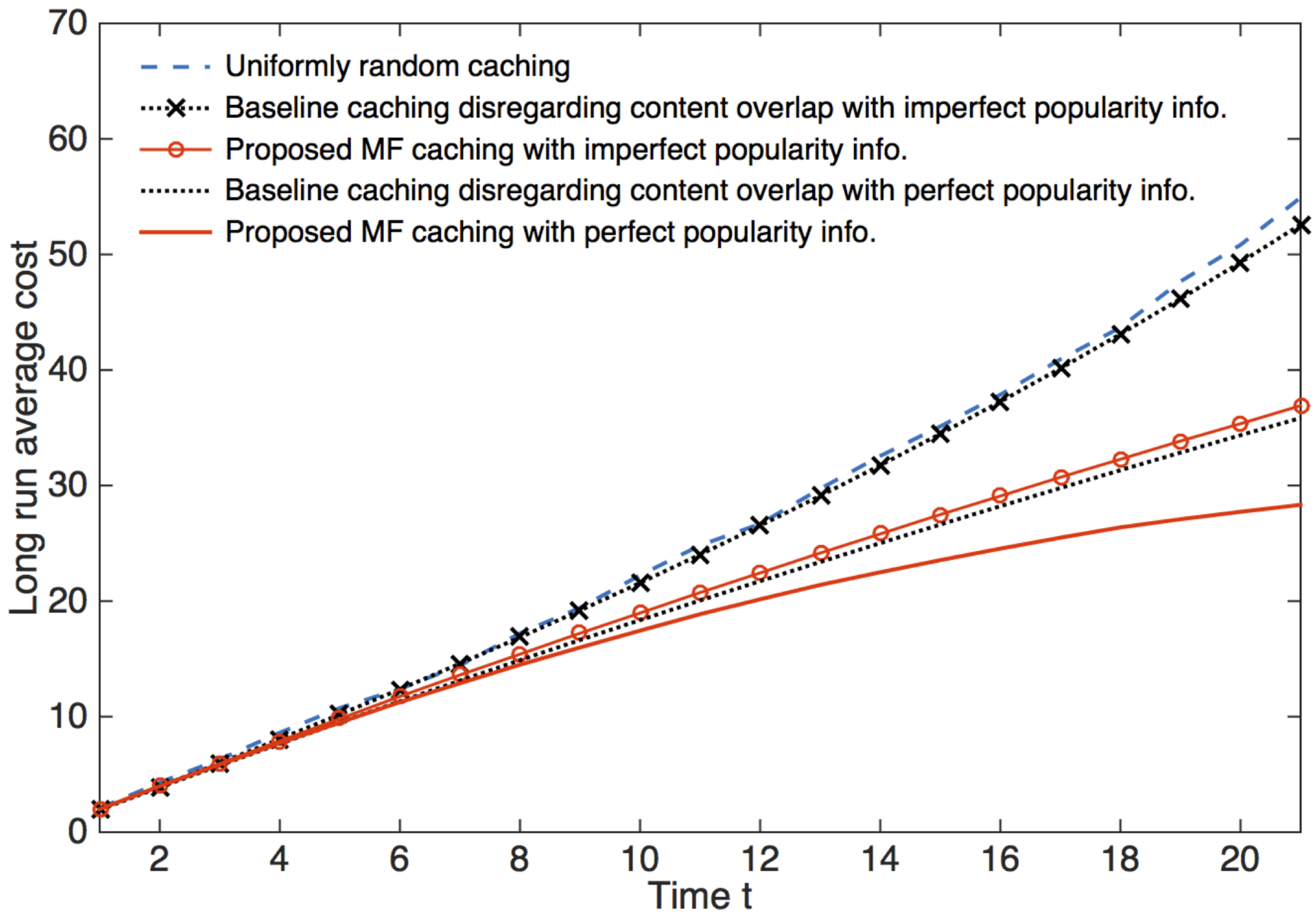}   
            \caption{{\small Long run average costs of different caching strategies with perfect and imperfect popularity information. ($Q(0)=0.7, x(0) =0.3, \eta=0.1$).} }\label{LRA} 
            \end{figure}

            This section evaluates the performance of the proposed MF caching algorithm under the spatio-temporal content popularity dynamics. 
            %
            Additionally, we evaluate the robustness of our scheme to imperfect popularity information in terms of the LRA caching cost.
            To this end, we compare the performance of the proposed MF caching algorithm with the following caching algorithms.
            \begin{itemize}
            \item {\it Baseline caching algorithm} that   does not consider the amount of content overlap but determines the instantaneous caching amount $\hat{p}_{j}(t)$ proportionally to the instantaneous request probability $x_j(t)$ subject to current backhaul, storage state, and interference described as follows: 
            $\hat{p}_{j}(t)=\frac{1}{L_j}\left[B_{j}(t)- \frac{1}{1+\mathcal{R}(t,I^f(t)) x_j(t)   } \right]^+$. 
            \item {\it Uniformly random caching} that randomly determines the caching amount following the uniform distribution.
            \end{itemize}

            \textbf{LRA Cost Comparison.}\quad Fig.~\ref{LRA_user} shows the LRA cost evaluation of the proposed MF caching algorithm, uniformly random caching, and the baseline caching  algorithm, which disregards the content overlap among neighboring SBSs.
             The LRA costs over time for different user density $\lambda_u$ are numerically evaluated. The proposed caching control algorithm reduces about $24\%$ of the LRA cost as compared to the caching algorithm without considering the content overlap. This performance gain is due to avoiding redundant content overlap and having an SBS under lower interference environment to cache more contents. As the user density $\lambda_u$ becomes higher for a fixed SBS density $\lambda_b$, the final values of the LRA cost increase for all the three caching schemes. When UDCNs are populated by numerous users, the fluctuation of  spatial dynamics of popularity increases and the number of  SBSs having associated users increases. 
            Hence, both the aggregate interference imposed by the SBSs and the content popularity severely change over the spatial domain. In this environment,  the advantage of the proposed algorithm compared to the popularity based algorithm becomes larger, yielding a  higher gap between the final values of the produced LRA cost.

            \begin{figure}
            \centering
            \includegraphics[angle=0, width=8cm]{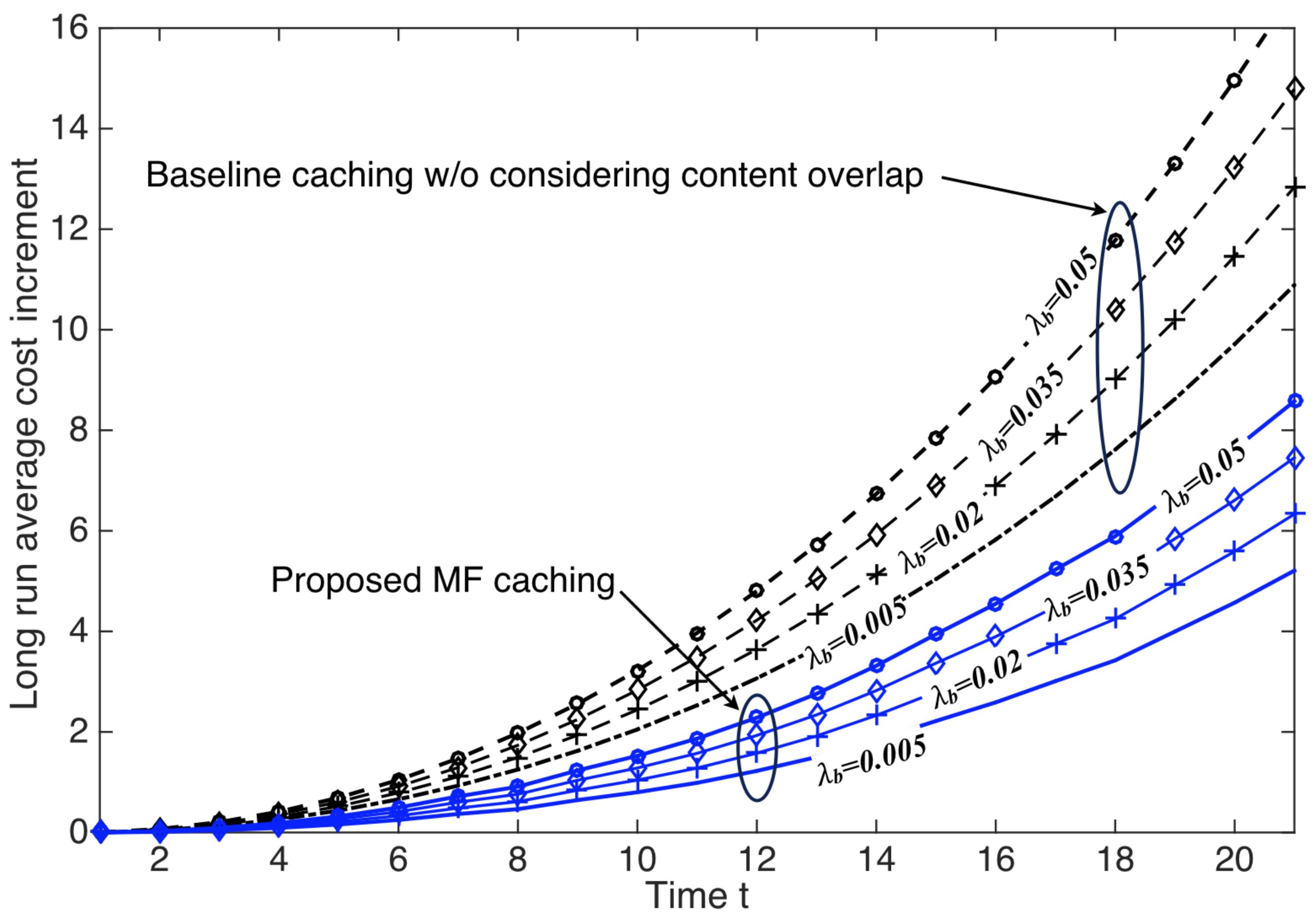}   
            \caption{{\small LRA cost Increment due to imperfect popularity information. For different SBS density  $\lambda_b$, the proposed MF caching and the baseline caching without considering the content overlap are compared ($Q(0)=0.7, x(0) =0.3, \eta=0.1$).} }\label{LRA_incre} 
            \end{figure}
            
            \begin{figure}
            \centering
            \includegraphics[width=.8\textwidth]{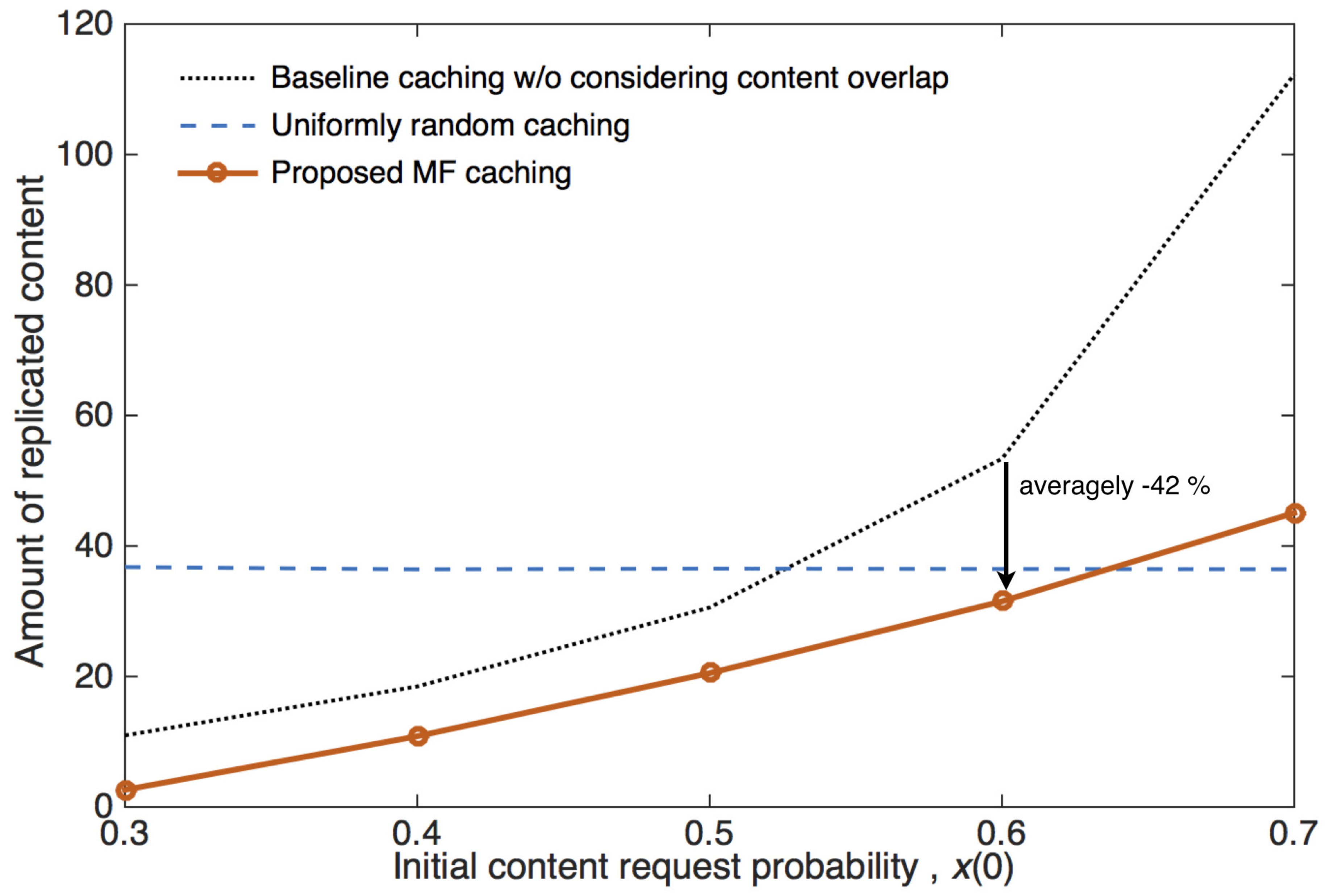}   
            \caption{{\small The amount of overlapping contents per storage usage ($Q(0)=0.7, \eta=0.1$). } }\label{Repli}
            \end{figure}
            
            \textbf{Demand Misprediction Impact.}\quad Accurate content popularity information may not be available at SBSs due to misprediction or estimation error of content popularity. It is thus assessed how the proposed algorithm is robust against imperfect popularity information (IPI) given as follows:
            \begin{align}
            \hat{x}(t)=x(t)+\Delta(t), \label{IPI_eq}
            \end{align}
            where $\hat{x}(t)$ denotes a content request probability estimated by an SBS, and $\Delta(t)$ represents an observation error for the request probability $x(t)$ at time $t$. An SBS has perfect popularity information (PPI) if $\Delta(t)$ is equal to zero for all $t$ (i.e. $\hat{x}(t)=x(t)$). The magnitude of $\Delta$ determines the accuracy of the popularity. 
             
            For numerical evaluations, an observation error $\Delta$ is assumed to follow a normal distribution $\mathcal{N}(0.2,0.001^2)$. SBSs respectively determine their own caching control strategies based on imperfect content request probability $\hat{x}(t)$ \eqref{IPI_eq} instead of PPI $x(t)$. With this IPI, the LRA caching cost over time is evaluated as shown in Fig. \ref{LRA}. The impact of IPI increases with the number of SBSs because redundant caching occurs at several SBSs.
            Also, the LRA increment due to IPI is evaluated for our MF caching algorithm and the popularity based one for different SBS density, i.e., the number of neighboring SBSs as shown in Fig. \ref{LRA_incre}. 
            The numerical results corroborate that the proposed algorithm is more robust against imperfect information of content popularity in comparison with the popularity-based benchmark scheme. In particular, our caching strategy reduces about $50\%$ of the LRA cost increment as compared to the popularity-based baseline method.

            Fig. \ref{Repli} shows the amount of overlapping contents per storage usage as a function of the initial content probability $x(0)$. The proposed MF caching algorithm reduces caching content overlap averagely 42\% compared to popularity based caching. However, MF caching algorithm yields a higher amount of content overlap than random caching does when the content request probability becomes high. The reason is that the random policy downloads contents regardless of their  popularity, so the amount of content overlap remains steady. On the other hand, MF caching increases the downloaded volume of popular content.

            \section{Conclusion}\label{Conclusion_remark}
            
            In this chapter, scalable and distributed edge caching in a UDCN has been investigated. To accurately reflect time-varying local content popularity, spatio-temporal content popularity modeling and interference analysis have been applied in optimizing the edge caching strategy. Finally, by leveraging MFG, the computing complexity of optimizing the caching strategy has been reduced to a constant overhead from the cost exponentially increasing with the number of SBSs in conventional methods. Numerical simulations corroborate that the proposed MFG-theoretic edge caching yields lower LRA costs while achieving more robustness against imperfect content popularity information, compared to several benchmark schemes ignoring content popularity fluctuations or cached content overlap among neighboring SBSs.

\bibliographystyle{vancouver-modified}

\end{document}